# High Thermoelectric Powerfactor in 2D Crystals of MoS₂

Kedar Hippalgaonkar[1, 2, 3, †], Ying Wang[1, †], Yu Ye[1, †], Diana Y. Qiu[2,4], Hanyu Zhu[1], Yuan Wang[1, 2], Joel Moore[2, 4], Steven G. Louie[2,4], Xiang Zhang[1, 2, *]

[1] NSF Nano-scale Science and Engineering Center (NSEC), 3112 Etcheverry Hall, University of California, Berkeley, California 94720, USA

[2] Material Sciences Division, Lawrence Berkeley National Laboratory (LBNL), 1 Cyclotron Road, Berkeley, CA 94720, USA

[3] Institute of Materials Research and Engineering, Agency for Science Technology and Research, Singapore, 117602

[4] Department of Physics, University of California, Berkeley, California 94720 USA

† These authors contributed equally.

*Correspondence and requests for materials should be addressed to X. Z. (email: xiang@berkeley.edu)

# ABSTRACT

The quest for high-efficiency heat-to-electricity conversion has been one of the major driving forces towards renewable energy production for the future. Efficient thermoelectric devices require high voltage generation from a temperature gradient and a large electrical conductivity, while maintaining a low thermal conductivity. For a given thermal conductivity and temperature, the thermoelectric powerfactor is determined by the electronic structure of the material. Low dimensionality (1D and 2D) opens new routes to high powerfactor due to the unique density of states (DOS) of confined electrons and holes. 2D transition metal dichalcogenide (TMDC) semiconductors represent a new class of thermoelectric materials not only due to such confinement effects, but especially due to their large effective masses and valley degeneracies. Here we report a powerfactor of $MoS_2$ as large as 8.5 $mWm^{-1}K^{-2}$ at room temperature, which is amongst the highest measured in traditional, gapped thermoelectric materials. To obtain these high powerfactors, we perform thermoelectric measurements on few-layer $MoS_2$ in the metallic regime, which allows us to access the 2D DOS near the conduction band edge and exploit the effect of 2D confinement on electron scattering rates, which result in a large Seebeck coefficient. The demonstrated high, electronically modulated powerfactor in 2D TMDCs holds promise for efficient thermoelectric energy conversion.

# I. INTRODUCTION

An ideal thermoelectric material behaves as an electron-crystal and phonon-glass, allowing a large temperature gradient across it while conducting electricity efficiently to generate a thermoelectric voltage [1]. Significant progress in the thermoelectric performance of materials has been made by exploring ultralow thermal conductivity at high temperature [2] and reducing thermal conductivity by nanostructuring [3], as well as by resonant doping [4] and energy-dependent scattering [5] of electrons. Recently, 2D transition metal dichalcogenides (TMDCs) have shown unique valley-dependent electronic and optical properties [6–10], and also have been theoretically predicted to be superior thermoelectric materials [11–13]. Most theoretical analyses are centered on low lattice thermal conductivity, but the latest calculations suggest that favorable electronic properties of TMDCs also result in an enhanced Seebeck effect [6,14–17], different from gapless, massless carriers in semi-metallic graphene [18–22]. Recent experiments have studied the photo-thermoelectric effect and Seebeck coefficient of monolayer $MoS_2$ at low carrier densities in the insulating regime, but low electrical conductivity limits its powerfactor for thermoelectric applications [23,24]. Here, we examine thermoelectric transport in 2D crystals of few-layer $MoS_2$ at high carrier concentrations in the metallic regime and observe powerfactors, $S^2\sigma$, as large as 8.5 $mWm^{-1}K^{-2}$ in bilayer $MoS_2$, where $S$ is the Seebeck coefficient and $\sigma$ is the electrical conductivity. We use the Seebeck coefficient to probe the 2D density of states (DOS) in both monolayer and bilayer $MoS_2$ and show that it agrees well with first-principles calculations. Moreover, we show that confinement effects on the electronic DOS and scattering rate enhance the Seebeck coefficient in 2D and the bilayer, in particular, has a larger value as a consequence of the higher effective mass and larger valley degeneracy. 2D TMDCs with high powerfactors are promising thermoelectric materials for planar applications such as Peltier cooling devices.

## II.    RESULTS

### A.  Gate-dependent powerfactor at room temperature

The Seebeck coefficient and electrical conductivity of 2D $MoS_2$ are measured as a function of carrier concentration tuned by a back gate (Fig. 1, see Methods for detailed measurement process). The electron concentration is given by $n = C_{ox}/e \cdot (V_g - V_t)$, where $C_{ox}$ is the capacitance between the channel and the back gate, $e$ is the electron charge, and $V_g$ and $V_t$ are the gate and threshold voltage, respectively. The measured electrical conductivities and Seebeck coefficients of monolayer, bilayer and trilayer $MoS_2$ follow behavior akin to an extrinsically doped semiconductor (Fig. 2a). The Seebeck voltage is proportional to the asymmetry of occupied density of states around the Fermi level [5,25]. Hence, with increasing electron concentration, the magnitude of the Seebeck coefficient drops as the Fermi level is pushed closer to the conduction band minimum (CBM) (see Methods for measurement details). However, the measured powerfactor $S^2\sigma$ increases correspondingly with applied gate voltage $V_g$ due to increasing electrical conductivity (Fig. 2b).  The bilayer device exhibits the largest powerfactor $S^2\sigma = 8.5$ $mWm^{-1}K^{-2}$ at $V_g = 104$ V equivalent to a high electron concentration of $n_{2D} \sim 1.06 \times 10^{13}$ $cm^{-2}$.

The magnitude of the powerfactor is expected to reach a peak and then drop for even higher carrier concentrations as the increasing electrical conductivity is offset by the decreasing Seebeck coefficient [5].  However, for our $MoS_2$ samples, the powerfactor does not peak, as this optimum carrier concentration is expected to occur at an even higher gate voltage ($n_{2D} \sim 1.31 \times 10^{13}$ $cm^{-2}$ equivalent to a bulk concentration of $n_{3D} \sim 1 \times 10^{20}$ $cm^{-3}$ – obtained by considering a bilayer thickness of 1.3 nm), which is limited by the electrical breakdown of the gate oxide in our experiment [26].

The effective mobilities are determined by a standard transistor measurement (see Supplemental Material for details). The measured effective mobilities at room temperature are 37 cm$^2$V$^{-1}$s$^{-1}$ for the monolayer and 64 cm$^2$V$^{-1}$s$^{-1}$ for the bilayer. We have measured a total of 11 monolayer and bilayer devices at room temperature and computed the powerfactor at a fixed carrier concentration, $n_{2D}$=5.2×10$^{12}$ cm$^{-2}$ (($V_g$−$V_t$)=66 V) (see Methods for details), which is lower than the highest carrier density where the maximum powerfactor is observed (see Supplemental Material for a summary of all measured devices) to establish the repeatability of our data. The bilayer sample shows the largest electrical conductivity as well as the highest Seebeck. Note that our samples are exfoliated from natural molybdenite crystals, so their initial dopant and impurity levels vary. Hence, the device mobilities differ from sample to sample and are lower than the theoretical estimate (~410 cm$^2$V$^{-1}$s$^{-1}$) [27], which could be due to extrinsic effects such as screening and scattering from the underlying dielectric substrate [16] and impurity levels in individual samples [28]. For phonon-limited theoretical mobility in suspended MoS$_2$, a powerfactor as large as 28 mWm$^{-1}$K$^{-2}$ is predicted at $n_{2D}$=1×10$^{12}$ cm$^{-2}$ [13]; therefore, in principle, the powerfactor of 2D MoS$_2$ can be improved further by making cleaner samples to obtain higher mobility closer to the theoretical limit.

### B. Temperature dependent transport in monolayer MoS$_2$:

At high temperatures and high electron concentrations, when the Fermi level is pushed close to the conduction band edge, monolayer MoS$_2$ undergoes an insulator-to-metal transition [14–16]. This metal-like regime for conducting MoS$_2$ is determined by analyzing the conductivity as a function of temperature for different electron concentrations (gate voltages): we study the temperature dependent electrical conductivity and Seebeck coefficient from 1.0×10$^{11}$ cm$^{-2}$ to 5.1×10$^{12}$ cm$^{-2}$ for a monolayer MoS$_2$ sample (see Supplemental

Material). The insulator-to-metal transition temperature ($T_{IMT}$) is defined as the temperature at which the measured conductivity changes from increasing with temperature to a metal-like decrease with temperature. We thus illustrate the electronic phase diagram of transport in MoS$_2$ (Fig. 3a) where $T_{IMT}$ is plotted as a function of the carrier concentration. In the insulating phase, the conductivity follows a relation in temperature given by: $\sigma \sqcup$ exp($-(T_0/T)^{1/3}$) in a 2D system, which fits a Mott Variable-Range-Hopping (m-VRH) model [16,28,29]. Concurrently, the measured Seebeck coefficient shows a monotonic increase with temperature as $S \sqcup T^{1/3}$ (Fig. 3b), using Zyvagin's formula for the m-VRH model [30–32], with $S \rightarrow 0$ as $T \rightarrow 0$ (inset of Fig. 3b). Similar m-VRH transport phenomenon has also recently been observed in CVD-grown MoS$_2$ for the insulating phase [23], in stark contrast with thermally activated transport mechanism in semiconductors [33,34]. Therefore, from the electronic phase diagram (Fig. 3a) for high temperatures ($T$>250 K) and large electron concentrations ($n > 2\times10^{12}$ cm$^{-2}$ at 300 K), electrical transport in MoS$_2$ is metal-like and the Mott relation for calculation of the Seebeck coefficient holds (see Supplemental Material for measured Seebeck at higher temperature for monolayer and bilayer devices). The doping level is not high enough to observe metallic transport behavior at lower temperatures.

### C. Nature of scattering in monolayer and bilayer MoS$_2$:

High powerfactors in 2D MoS$_2$ have been predicted to stem from large conduction band effective masses, leading to a large Seebeck coefficient [13]. In order to better understand the origin of the large Seebeck magnitude for monolayer and bilayer MoS$_2$, we calculate the Seebeck from the linearized Boltzmann Transport Equation (BTE) under the relaxation time approximation, given by:

$$S = \frac{1}{qT}\left[\frac{\int_{Ec}^{\infty}\frac{df_{FD}}{dE}D_{2D}\left(E\right)\left(E - E_F\right)t\left(E\right)dE}{\int_{Ec}^{\infty}\frac{df_{FD}}{dE}D_{2D}\left(E\right)E\,t\left(E\right)dE}\right] \qquad (1)$$

Here, $f_{FD}$ is the Fermi Dirac distribution, $D_{2D}(E)$ is the 2D density of states, $E_F$ is the Fermi Level with respect to the CBM at $E_c$, $q$ is the electron charge, and $\tau(E) = \tau_0 E^r$ is the energy-dependent relaxation time (details in Supplemental Material), where $r$ is the scattering exponent and depends on the dominant scattering mechanism. In order to obtain the density of states used in Equation (1) above, we performed first-principles calculations of the quasiparticle (QP) bandstructure of suspended monolayer and bilayer $MoS_2$ within the GW approximation [35] (details in Supplemental Material). The conduction band minimum was found to be at the K and K' points in the Brillouin zone for monolayer $MoS_2$ and along the six-fold degenerate Λ-high-symmetry line (Λ valley) for bilayer $MoS_2$, in good agreement with previous calculations [36–38]. The computed DOS of pristine monolayer and bilayer $MoS_2$ at the GW level show that due to the larger band effective mass and higher degeneracy in the Λ-valley, the DOS of bilayer $MoS_2$ at the CBM is ~4 times larger than the DOS of monolayer $MoS_2$ (details in Supplemental Material). As expected for parabolic bands in 2D, we observe that the DOS is a step function at the conduction band edge in both cases (Figs. 4a and 4b: The broadening seen in the figures results from a numerical 20meV broadening in the calculation). Thus, in estimating the Seebeck from equation (1) above, we assume that the DOS is constant, given by the value of the DOS at the step edge (dotted vertical lines in Figs. 4a and 4b) and hence energy-independent.

The parameters derived from the band structure calculations that are used for determining the Seebeck coefficient are given in Table 1. We can solve for the position of the fermi level, $E_F$ with respect to the CBM, $E_c$ as a function of the backgate voltage, which is linked to the induced carrier concentration in the $MoS_2$ channel (Supplemental Material). Then, we calculate the Seebeck coefficient for both monolayer and bilayer $MoS_2$ as a function of the carrier concentration, and compare the calculated Seebeck to experimental values for four different devices each (Figs. 4c and d respectively). Here, we see that the Seebeck, as calculated from Equation (1), fits the experimental data quite well when r=0, which is consistent with phonon-limited scattering in 2D (see Supplemental Information), and captures the relative change in the Seebeck as a function of the carrier concentration induced by the backgate voltage. Finally, our calculations show that given identical carrier concentrations, the magnitude of the Seebeck for the bilayer is larger than that for the monolayer, as a consequence of the larger density of states at the conduction band edge, which stems from both the heavier effective mass as well as a higher valley degeneracy of the CBM at the high symmetry Λ-valley.

## III.    DISCUSSION

The Seebeck coefficient is given by integrating the energy dependent relaxation time modulated by a window function defined by $F_{wI}(E,T) = (E-E_F) \times \{-df_{FD}(E,T)/dE\}$, where $E_F$ is the Fermi level and $f_{FD}(E,T)$ is the Fermi-Dirac distribution [39]. This function is odd around $E_F$, with a width of $\sim 2k_BT$ [40]. For doped, metal-like monolayer MoS$_2$, as the Fermi level approaches the bottom of the conduction band within this energy width of the window function, the rapidly changing DOS (Fig. 4a and 4c) generates a large asymmetry around the Fermi level, which leads to an enhanced value of the Seebeck coefficient [41,42]. This effect is exacerbated by the large transport effective mass ($m_d^*$), which includes the valley and spin degeneracies. In three dimensions, $m_d^* = (g_v \cdot g_s)^{2/3} \times m^*$ [1,40]. In two dimensions, $m_d^* = (g_v \cdot g_s) \times m^*$; for monolayer MoS$_2$, $g_v = g_s = 2$, and thus the density of states effective mass contributing to transport is $m_{d,1L}^* \sim 2.1m_0$. Bilayer MoS$_2$ has $g_v = 6$ and $g_s = 2$, giving $m_{d,2L}^* \sim 8.1m_0$. These values are significantly larger than conventional thermoelectric materials and indeed are the main reason for our large measured Seebeck coefficients.

The fits to the Seebeck coefficient in Figs 4c and 4d using the full Fermi-Dirac distributions are accurate for carrier concentrations higher than n $\sim$2-4x10$^{12}$ cm$^{-2}$, which is consistent with the phase diagram in Fig. 3a. At lower temperatures and lower carrier concentrations, VRH transport is determined by a localization length up to n $\sim$ 2x10$^{12}$ cm$^{-2}$ (Supplemental Material). We have also considered bandgap renormalization in monolayer MoS$_2$ at high doping levels ($\sim$1x10$^{13}$ cm$^{-2}$) and establish that the average effective mass decreases slightly with doping concentration, thus explaining the slight drop in the measured Seebeck coefficient at high carrier concentrations in Fig. 4c (details in Supplemental Material). Notwithstanding these minor effects, the scattering exponent ($r = 0$) determined from fitting the calculated Seebeck coefficients to the experimental data (Figs. 4c and 4d) as well as the exponent of the temperature-dependent mobility, $m \sim T^{-1.9}$ at high temperatures (Supplemental Material) prove that transport in supported, doped 2D MoS$_2$ (and probably more generally in TMDCs) is limited by phonon scattering at high temperatures.

Despite the excellent agreement of experimental and theoretical Seebeck coefficient, our measured field-effect mobility is still much lower than the calculated, intrinsic value of 410 cm$^2$/V.s [27] because in the calculation of the intrinsic mobility the total scattering rate is obtained as a sum over all the phonon channels only in pristine MoS$_2$. It's not surprising, in our case, that the substrate would add additional scattering channels, thus reducing the mobility further. Indeed, our measured values of mobility ($\sim$10-60 cm$^2$/V.s) are comparable to other experiments on 2D MoS$_2$/SiO$_2$ [6,17,43]. Intriguingly, as the Seebeck coefficient does not depend on the energy-independent magnitude of the scattering time, $\tau_0$, but instead only on the energy-dependent exponent, $r$, there are many avenues to improve the measured powerfactor further by judiciously picking substrates with weak phonon-coupling, as well as improving the quality of the MoS$_2$ channel. The magnitude of the Seebeck coefficient is expected to be even larger when the relaxation time has energy-dependence with r>0 (r=1.5 is plotted for reference in Figs. 4c and 4d), so engineering the dielectric environment to change the dominant scattering mechanism is another possible route to enhance the powerfactor. Like $MoS_2$, other TMDCs [44] and phosphorene [45,46] are expected to simultaneously have large band effective masses and mobilities possibly leading to high values of powerfactor, thus highlighting 2D semiconductor crystals as potential thermoelectric materials. It remains to be seen if the thermal conductivity of these materials can be tuned further, making them directly useful for thermoelectric applications by enhancing the thermoelectric figure-of-merit ZT, although a high powerfactor itself can be utilized for in-plane Peltier cooling [47].

## IV.    CONCLUSIONS

Our experiments report the thermoelectric properties of exfoliated 2D crystals of $MoS_2$, and we observe high powerfactors as large as 8.5 $mWm^{-1}K^{-2}$ at room temperature. This is twice as high as commonly used bulk $Bi_2Te_3$, making 2D TMDCs promising candidates for planar thermoelectric applications. The enhanced powerfactor in the metallic regime is attributed to the sizable conductivity in the highly doped crystals and a large Seebeck coefficient stemming from high valley degeneracies and effective masses, especially in the case of the bilayer where a large effective mass at the CBM in the $\Lambda$-valley is coupled with a 6-fold valley degeneracy. We measure thermoelectric transport in the highly doped regime for the first time, thus allowing us to access the 2D density of states in TMDCs. Our device configuration allows us to tune the carrier concentration of 2D $MoS_2$, which is difficult in bulk materials, hence providing important insights into thermoelectric transport in these layered materials. The high powerfactor in layered TMDCs provides an exciting avenue to enhance thermoelectric efficiencies and galvanize the growth of thermoelectric devices in the near future.

**FIGURES**

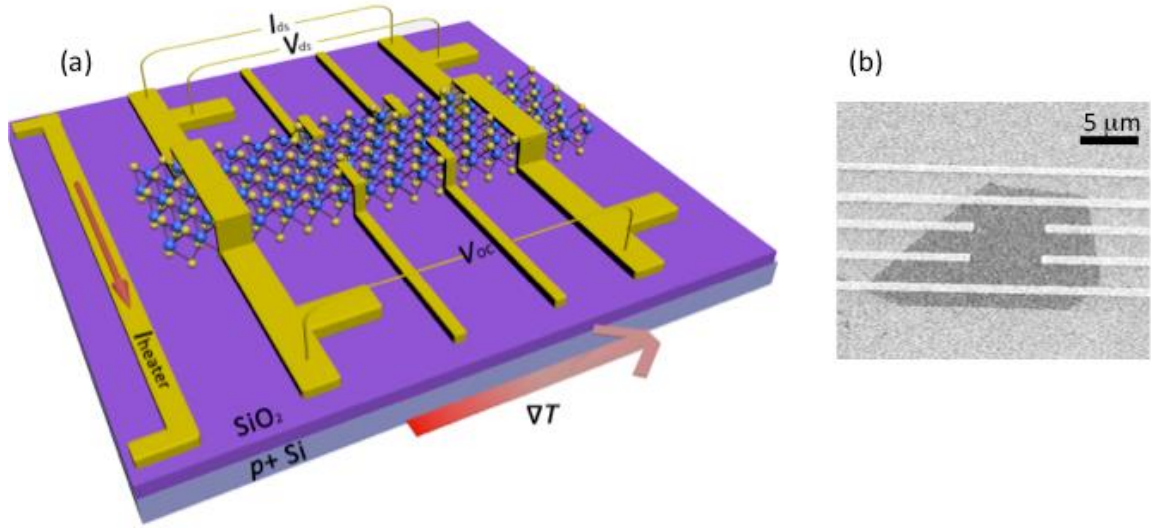

**Figure 1: Device Configuration for Seebeck coefficient and electrical conductivity measurement of MoS₂** (a) Schematic of the simultaneous measurement of the Seebeck coefficient and the electrical conductivity. The illustration shows a monolayer MoS₂, placed on thermally grown SiO₂ on a *p*+ silicon substrate. Two-probe electrical conductivity was measured by passing a current through the device ($I_{ds}$) and measuring the drain-source voltage ($V_{ds}$) at each temperature. In order to measure the Seebeck coefficient $S = -V_{oc}/\Delta T$, current was passed through the heater to generate a temperature gradient, $\Delta T$ while the open circuit voltage ($V_{oc}$) was measured. (b) Scanning electron micrograph of an actual device as described in (a). Note, the hall-bar electrodes were used to obtain the ratio of the two-probe to the four-probe electrical conductivities, $\gamma_c = \sigma_{4p}/\sigma_{2p}$ to estimate the contribution due to contact resistance at each temperature. For the monolayer sample, $\gamma_c = 1.98$ at 300 K with temperature dependent ratios shown in the Supplemental Material.

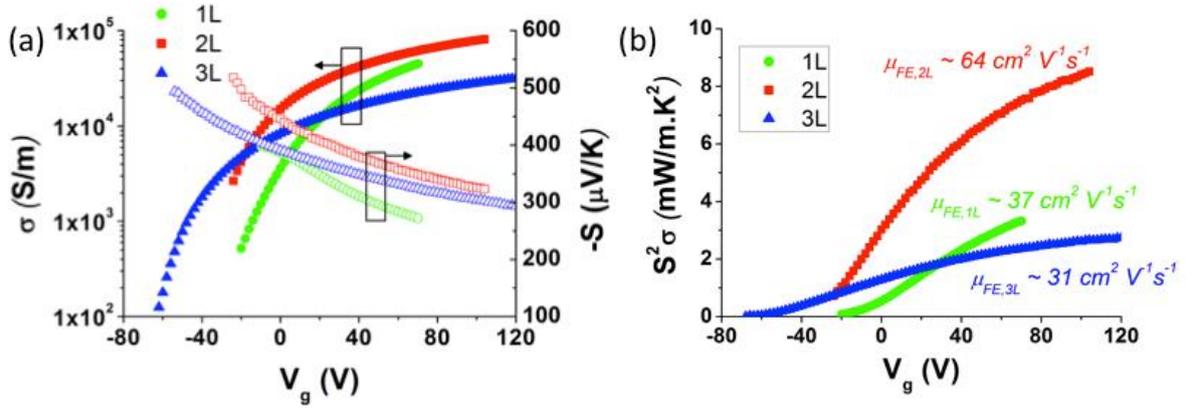

**Figure 2: Thermoelectric characterization of 2D MoS₂ at room temperature.** (a) Electrical conductivities, (closed markers) and Seebeck coefficients, $S$ (open markers) as a function of gate voltage at 300 K for monolayer (green circles), bilayer (red squares) and trilayer MoS₂ (blue triangles). As the carrier concentration $n \sqcup (V_g - V_t)$ increases, $\sigma$ increases and the magnitude of $S$ decreases. $S$ is negative, which confirms that the sample is $n$-type. (b) Powerfactor, $S^2\sigma$ as a function of $V_g$. The bilayer device with a larger effective mobility of 64 cm²V⁻¹s⁻¹ exhibits maximum powerfactor of 8.5 mWm⁻¹K⁻² at $n$=1.06×10¹³ cm⁻² at room temperature, twice that of commercially used bulk Bi₂Te₃.

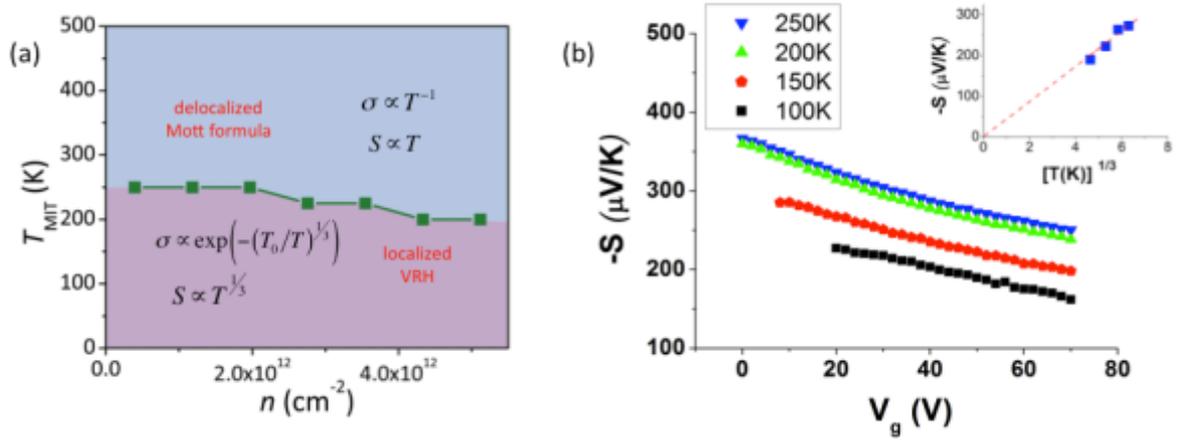

**Figure 3. Electronic phase transition and temperature dependent Seebeck in monolayer MoS₂.** (a) Phase diagram for thermoelectric transport as a function of temperature and electron concentration. For the metallic phase, $T > T_{IMT}$, electrical conductivity decreases with temperature, $\sigma \sqcup T^{-1}$ and the Seebeck coefficient increases slowly, $S \sqcup T$ (Mott formula for extended states). In the insulating phase, $T < T_{IMT}$, Mott-Variable Range Hopping for localized states dictates transport resulting in $\sigma \sqcup \exp(-(T_0/T)^{1/3})$ (see Supplemental Material) and $S \sqcup T^{1/3}$. (b) Experimental Seebeck coefficient for monolayer MoS₂ as a function of temperature and applied back-gate voltage. The magnitude of Seebeck decreases (increases) with $V_g$ (temperature). In the inset we show measured Seebeck at a fixed carrier concentration $n = C_{ox}/e \cdot (V_g - V_t)$, which follows a function of $T^{1/3}$, indicating m-VRH (localized) regime in the temperature range 100-250 K. At all temperatures, the experimental Seebeck at a fixed carrier concentration ($V_g - V_t$) is considered.

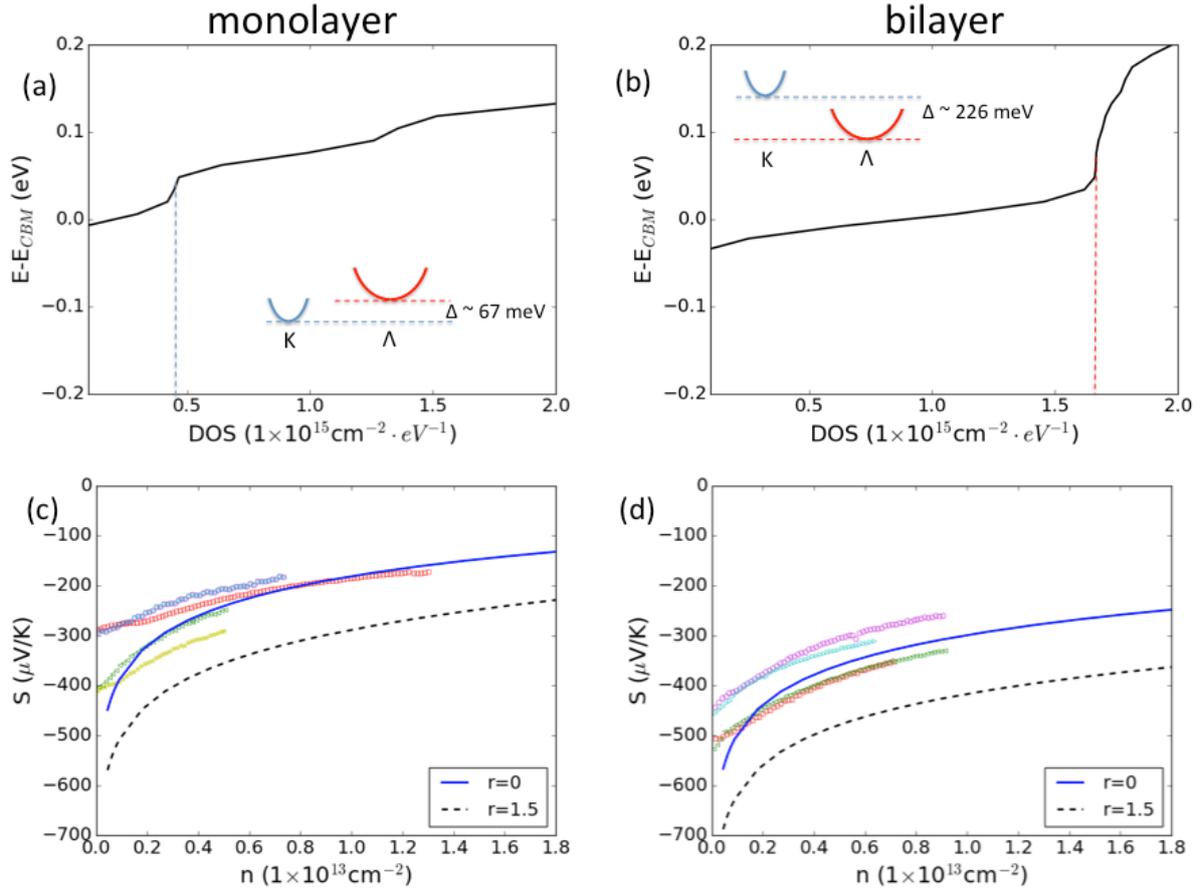

**Figure 4. Electronic Density of States (DOS) and carrier concentration dependent thermopower at 300K for monolayer and bilayer MoS₂.** Calculated DOS of pristine (a) monolayer and (b) bilayer MoS₂ plotted as a function of the energy difference from the conduction band minimum (CBM) in the K(Λ)-valley for the mono(bi)layer. The step function feature expected from 2D confinement can be seen clearly and is used to estimate the constant DOS (dotted vertical lines) used in Eq. (1). [inset: the relative positions of the K and Λ valleys in monolayer and bilayer MoS₂ showing that thermoelectric transport only occurs through the K-point in the monolayer and only through the Λ-high-symmetry direction in the bilayer, since the energy difference >∼2$k_BT$ in both cases]. (c) Monolayer and (d) bilayer experimental data (open symbols) compared with the estimated Seebeck coefficient from Eq. (1) for r=0, consistent with phonon-limited scattering in 2D, (solid lines) and r=1.5, for reference (dashed lines) − the data fits the r=0 phonon-limited scattering case well.

**Table I**. Summary of band structure calculations obtained from pristine monolayer and bilayer $MoS_2$ used for the estimation of the Seebeck coefficient using Eq. (1):

| | monolayer | bilayer |
|---|---|---|
| Valley Degeneracy: $g_v$ | 2 | 6 |
| Spin Degeneracy: $g_s$ | 2 | 2 |
| Effective Mass: $m*$ | (0.45+0.59)/2 $m_0$ ~ 0.52$m_0$ <br><br> at the K-point CBM | 0.68$m_0$ <br><br> at the $\Lambda$-point CBM |
| Density of States, $D_{2D}$ | 4.33x$10^{14}$ $cm^{-2}$ $eV^{-1}$ | 17.0x$10^{14}$ $cm^{-2}$ $eV^{-1}$ |

# APPENDIX A
## Sample preparation and characterization and measurement details

Exfoliated samples are obtained using the scotch-tape method by cleaving bulk molybdenite. We exfoliate the samples onto 275 nm thermally grown $SiO_2$ on a highly doped $p$-Si substrates. $MoS_2$ flakes are visible on the sample under an optical microscope and the monolayer, bilayer or trilayer samples are selected based on characterization using optical contrast, photoluminescence imaging and Raman Spectroscopy (see Supplemental Material). Layer thicknesses for monolayer and bilayer devices are measured with Atomic Force Microscopy (AFM) for fabricated samples. Defective samples with cracks, ripples and/or folds are identified with High-Resolution Scanning Electron Microscopy (HR-SEM) and are not used for measurements (see Supplemental Material).

The heating element is a resistive metal line, through which a DC current, $I_{DC}$, up to 20 mA is applied. The heat generated from the heater line creates a temperature gradient across the TMDC sample, given by $Q \sqcup I^2_{DC} R_{htr} \sqcup \Delta T$. The electrodes patterned on two sides of the sample function both as probes for electrical measurements and for local temperature measurement. For each electrode, the resistance is given by $R_{hot/cold} \sqcup T_{hot/cold}$. Then, the temperature difference across the device is calibrated as $\Delta T = T_{hot} - T_{cold}$, where $R_{hot/cold} = \alpha_{hot/cold} / T_{holt/cold}$ obtained at every global temperature, where the slope $\alpha_{hot/cold}$ is determined experimentally. The open circuit voltage across the device, $V_{oc}$, as a function of heating current is then determined, from which the Seebeck coefficient of the device can be deduced as $S = -V_{oc}/\Delta T$.

In order to minimize the electrical contact resistance, we use Ti/Au films evaporated with electron beam evaporation. Titanium has been known to have good Fermi level alignment with monolayer $MoS_2$ [48]. In order to improve the contact quality, we annealed the sample *in-situ* at 475 K for one hour in the cryostat prior to performing measurements. After annealing, all of our *I-V* curves are linear, indicating ohmic contact and hence none of the transport characteristics can be ascribed to Schottky behavior. We have characterized the contact resistance of the Ti/Au contacts and found the ratio of four-probe to two-probe conductance at all temperatures (see Supplemental Material). It has been reported that the contact resistance contribution to measured total resistance at room temperature can be as large as 50% at 100 K with Ti/Au contacts [48]. In our case, we define the ratio of the four-probe to the two-probe conductivity as the contact ratio, $\gamma_c$, which is 2 at 300 K and 2.5 at 100 K. Hence, our estimation of the intrinsic electrical conductivity of the layered $MoS_2$ is underestimated due to included contact resistance. The Seebeck measurements are not affected by the contact quality since they are measured in an open-circuit configuration. However, the measured *S* is a sum of the sample and the contacts (Ti/Au). Since the metallic Seebeck is < 1 $\mu VK^{-1}$ at all temperatures, it does not affect our measurements and we do not consider it in our estimation. The effects of joule heating, current crowding and thermoelectric potentials due to current flow in the 2D devices[41, 42] is negligible since the current densities used for electrical conductivity measurements are very small, $I_{ds}$ < 0.1 $\mu$A/$\mu$m (see Supplemental Material). All measurements were performed in vacuum at $2\times10^{-6}$ torr. For lower gate voltages close to the threshold voltage $V_t$, the channel resistance becomes too high and we are unable to measure the Seebeck coefficient accurately. The maximum gate voltages $V_g$ applied for all devices are limited by the electrical breakdown of the gate oxide. In order to determine identical carrier concentrations (*n*) for different devices, we determined the threshold voltage ($V_t$) by linear extrapolation of the transfer curve ($I_{ds}$ vs $V_g$). Since each device has a different $V_t$, the gate voltage at which the powerfactor is considered (for same carrier concentration) is also different for each device.

# Supplementary Material for High Thermoelectric Powerfactor in 2D Crystals of MoS₂

Kedar Hippalgaonkar[1, 2, 3, †], Ying Wang[1, †], Yu Ye[1, †], Diana Y. Qiu[2,4], Hanyu Zhu[1], Yuan Wang[1, 2], Joel Moore[2, 4], Steven G. Louie[2,4], Xiang Zhang[1, 2, *]

[1] NSF Nano-scale Science and Engineering Center (NSEC), 3112 Etcheverry Hall, University of California, Berkeley, California 94720, USA

[2] Material Sciences Division, Lawrence Berkeley National Laboratory (LBNL), 1 Cyclotron Road, Berkeley, CA 94720, USA

[3] Institute of Materials Research and Engineering, Agency for Science Technology and Research, Singapore, 117602

[4] Department of Physics, University of California, Berkeley, California 94720 USA

† These authors contributed equally.

*Correspondence and requests for materials should be addressed to X. Z. (email: xiang@berkeley.edu)

# Calculating effective mass and density of states in monolayer and bilayer MoS₂:

The theoretical band structure and density of states calculations as described in the main text were done in a supercell arrangement with a plane-wave basis using norm-conserving pseudopotentials with a 125 Ry wave function cutoff. We included the Mo semicore 4d, 4p and 4s states as valence states for our DFT and GW calculations. The distance between repeated supercells in the out-of-plane direction was 25 Å. We fully relaxed the monolayer and bilayer MoS₂ structures and included spin-orbit interactions as a perturbation [1,2]. The dielectric matrix was calculated on a 60x60x1 q-point grid with a 25 Ry energy cutoff. 2500 bands were included in the summation over empty states. Dynamical effects in the screening were included with the Hybertsen-Louie generalized plasmon pole model (HL-GPP) [3].

The calculated QP bandstructures and density of states of monolayer and bilayer MoS₂ are shown in Supplemental Fig. 1 and Fig. 4a-b in the main text. We find that monolayer MoS₂ has a direct bandgap at the K point. In addition to the conduction band minimum (CBM) at K, there is another valley in the conduction band along the Λ-high-symmetry line from Γ to K. We find that the bottom of this Λ valley is 67 meV higher in energy than the K-point and thus unlikely to contribute to the Seebeck at room temperature. We find that spin-orbit coupling splits the conduction band at K by 2 meV, so we expect that both spin bands will contribute to the transport. We further determine that the effective mass of the lower band (which we will refer to as spin up) is $0.45m_0$, and the effective mass of the upper band (which we will refer to as spin down) is $0.59m_0$, where $m_0$ is the free electron mass. For bilayer MoS₂, we find that the CBM occurs along the Λ high-symmetry line. This Λ valley is anisotropic, and its average effective mass is $0.68m_0$. Calculated effective masses, spin-orbit (SO) splitting of the conduction band, and ordering of the conduction band valleys are summarized in Table I.

Finally, we explore the possibility that carrier doping, which is known to renormalize the QP band gap, might also change the QP effective masses. We performed an additional GW calculation on doped monolayer MoS₂, with a carrier concentration of $n=1 \times 10^{13}$ cm⁻². We found that QP effective mass of the spin up band in the K valley is unchanged for the spin-up band, while the effective mass of the spin down band decreases by 10%. Thus, the average carrier effective mass decrease by ~$0.08 \ m_0$ as the doping is increased from 0 to $n=1 \times 10^{13}$ cm⁻².

**Table I:** Comparison of 1. Difference between the conduction band minimum at K and along the Λ high-symmetry line ($E_K$-$E_\Lambda$), 2. SO splitting of the conduction band at K, and 3. effective masses for spin up (↑) and spin down (↓) states in the K and Λ valleys in units of the free electron mass ($m_0$) for monolayer and bilayer MoS₂ with different doping levels (n).

|  | n (cm$^{-2}$) | $E_K - E_\Lambda$ (eV) | $E_{K,c\downarrow} - E_{K,c\uparrow}$ (eV) | $m_{K\uparrow}$ ($m_0$) | $m_{K\downarrow}$ ($m_0$) | $m_{\Lambda\uparrow}$ ($m_0$) | $m_{\Lambda\downarrow}$ ($m_0$) |
|---|---|---|---|---|---|---|---|
| monolayer | 0 | -0.067 | 0.003 | 0.45 | 0.59 | 0.87 | 0.73 |
| monolayer | $1 \times 10^{13}$ | -0.668 | 0.003 | 0.45 | 0.53 | 1.18 | 1.02 |
| bilayer | 0 | 0.226 | 0.000 | 0.67 | 0.67 | 0.68 | 0.68 |

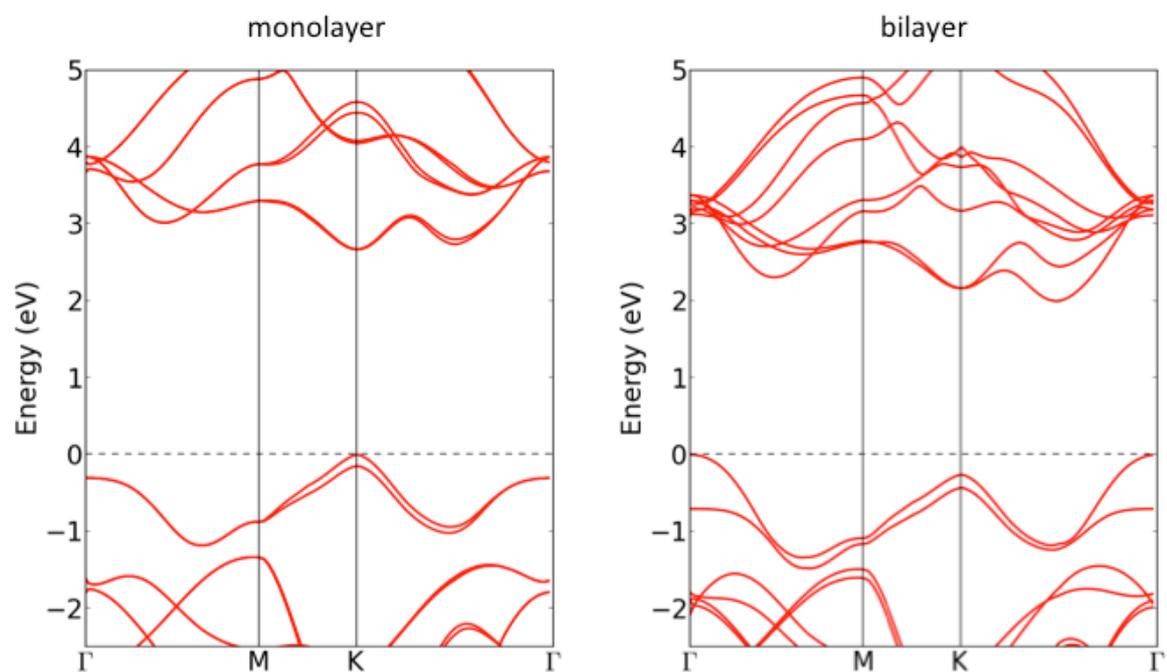

**Figure S1.** Quasiparticle bandstructure of pristine monolayer (left) and bilayer (right) MoS$_2$ calculated at the GW level

**Field-effect Mobility:**

The mobility of MoS$_2$ is determined following standard procedure using field-effect transistor with a fixed drain-source voltage, $V_{ds}$=10 mV. Here, the drain-source current, $I_{ds} \propto V_g$ and the mobility is given by: $\mu = \dfrac{dI_{ds}}{dV_g} \dfrac{L}{W} \dfrac{1}{V_{ds}} \dfrac{1}{C_{ox}}$ where $L$ and $W$ are the MoS$_2$ channel length and width respectively, and $C_{ox} = \varepsilon_r \varepsilon_0 / t_{ox} = 1.26 \times 10^{-4}$ F/m$^2$ is the oxide capacitance, where $\varepsilon_r$ = 3.9 is the relative permittivity of SiO$_2$, $\varepsilon_0$ = 8.85x10$^{-12}$ F/m is the permittivity of free space and $t_{ox}$ =275 nm is the thickness of the thermally grown oxide. This results in an estimated field mobility of 37 cm$^2$V$^{-1}$s$^{-1}$ for the monolayer, 64 cm$^2$V$^{-1}$s$^{-1}$ for the bilayer and 31 cm$^2$V$^{-1}$s$^{-1}$ for the trilayer [4,5]. Since the exfoliated MoS$_2$ are *n*-type semiconductors, the devices turn on at negative gate voltages; the turn-on voltage is −20 V for the monolayer, −38 V for the bilayer and −64 V for the trilayer. The range of mobilities is consistent with previous reports of high mobility between 1-100 cm$^2$V$^{-1}$s$^{-1}$. [4,6] Here we only report the two-probe mobilities for all devices.

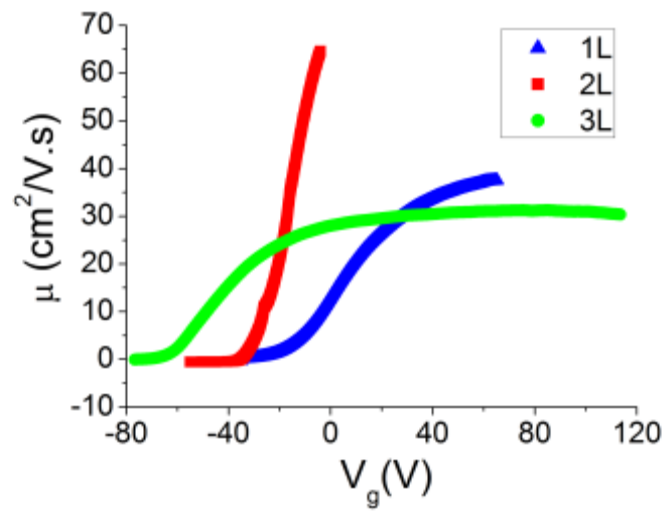

**Figure S2.** The measured field-effect mobilities of monolayer, bilayer and trilayer as a function of back gate ($V_g$). The measured mobility is 37 cm$^2$V$^{-1}$s$^{-1}$ for the monolayer, 64 cm$^2$V$^{-1}$s$^{-1}$ for the bilayer and 31 cm$^2$V$^{-1}$s$^{-1}$ for the trilayer.

**Summary of thermoelectric powerfactor measurements of monolayer and bilayer devices at 300K:**

We have measured 11 total devices with a range of maximum powerfactors (at maximum applied gate voltages) as seen in Fig. S3 below. Note here that if the back gate had not failed, we could have achieved even higher powerfactors for these devices. Since the mother crystals used for exfoliation are obtained from naturally mined sources, the initial level of doping (unknown) as well as the cleanliness/impurities of the particular peeled sample determines the mobility of the measured device and produces devices with a range of powerfactors as illustrated. 3 out of the 11 devices measured (PF= 8.5±1.5 mW/m.$K^2$, 5.0±1.0 mW/m.$K^2$ and 4.0±1.0 mW/m.$K^2$) have a powerfactor that is larger than commercial $Bi_2Te_3$, demonstrating repeatability of our high values.

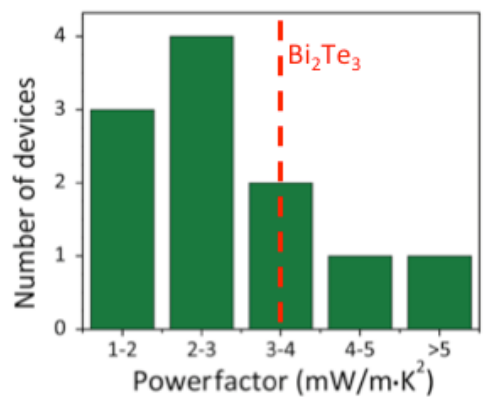

**Figure S3.** 11 monolayer and bilayer $MoS_2$ devices and their corresponding maximum powerfactor (at highest applied gate voltage) represented as green columns. 3 out of the 11 samples have a powerfactor ≥4 mW/m·$K^2$, which is larger than commercial $Bi_2Te_3$ (3.5±1.5 mW/m·$K^2$).

**Mott Variable Range Hopping (M-VRH) for insulating phase of monolayer MoS₂:**

Temperature and gate-voltage dependent conductivity is used to ascertain the Insulator-to-Metal Transition Temperature ($T_{IMT}$) (Fig. S4a). A metal-like behaviour is observed for $T>T_{MIT}$, as seen in the main text since $S \propto 1/T$ and is further corroborated by a rapid decrease of mobility as a function of temperature $\mu \propto T^{-1.9}$ (Fig. S4b). In the insulating phase for $T<T_{MIT}$, it is seen that the exponent is ~0.23, which is close to the expected behaviour of $m \propto T^{-1/3}$ in two dimensions [7]. Further, the conductivity in the insulating phase follows $\sigma(T) = \sigma_0 \exp[-(T_0/T)^{1/3}]$ in two dimensions (Fig. S4c) [7,8], and matches the Mott Variable Range Hopping (M-VRH) mechanism, where $T_0$ is related to the correlation energy scale. Here, electrons at the Fermi level below the mobility edge (at $E_c$) are localized, but are able to hop from one localized site to another due to the gradient field or interaction with phonons [8,9].

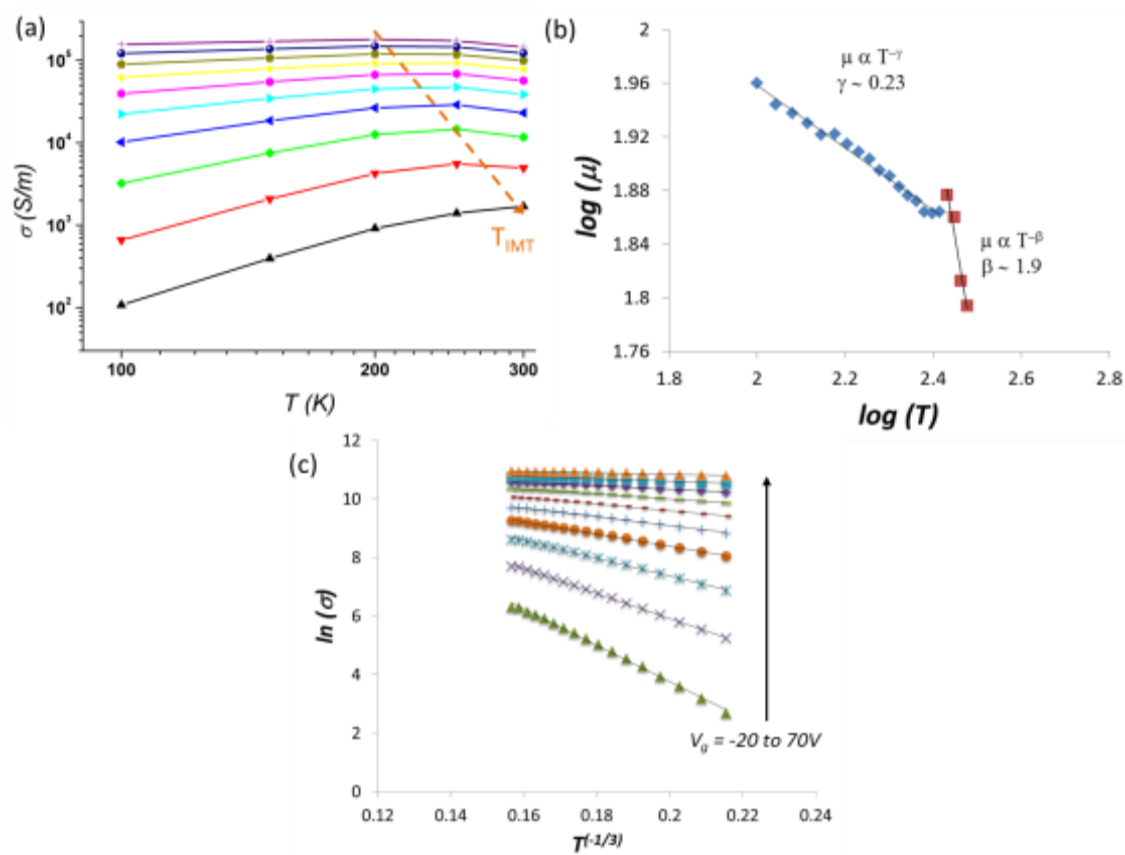

**Figure S4.** (a) Conductivity as a function of temperature (high ($n = 5 \times 10^{12}$ cm⁻² at the top) to low ($n = 4 \times 10^{11}$ cm⁻² at the bottom) carrier concentration). As the gate voltage (carrier concentration) decreases, the insulator-to-metal transition temperature ($T_{IMT}$) shifts to higher temperatures (indicated by the dotted arrow). (a) Temperature-dependent mobility of monolayer MoS₂. The mobility undergoes a rapid decrease with an exponent ~0.23 to ~1.9 crossing the metal-insulator-transition temperature ($T_{MIT}$). (b) Temperature-dependent conductance of monolayer MoS₂ in insulating phase with different gate voltages (carrier concentration). The conductance follows the relation of $\sigma(T) = \sigma_0 \exp[-(T_0/T)^{1/3}]$ in two dimensions, indicating the Mott Variable Range Hopping (M-VRH) mechanism.

Zhu et. al. have performed CV experiments on monolayer $MoS_2$ to obtain the electronic density of states due to the band as well as trapped states [10]. Their results indicate localized states with a density of (B-traps, $D_{it}$) ~2 − 8 x$10^{13}$ $eV^{-1}cm^{-2}$, while they assume a constant metallic density of states ~3.3x$10^{14}$ $eV^{-1}cm^{-2}$ ($m^* \sim 0.4m_0$).

This transition can be understood further by considering the localization length of hopping electrons in the insulating phase. In 2D, Mott VRH gives $T_0 = (13.8/k_B D(E_F)\xi^2)$ [11,12], where $D(E_F)$ is the density of states due to the localized states at the Fermi level, $E_F$ and $\xi$ is the localization length. If we assume the density of localized states is equal to that observed in Ref. 10, $D(E_F) = D_{it} \sim 8x10^{13}$ $eV^{-1}cm^{-2}$, we can extract the localization length, $\xi$ as a function of $T_0$ as shown in Figure S5(b) from our experiment on a monolayer $MoS_2$ sample saturating to a value of $\xi \sim 2.7nm$ when transport is fully metallic. This value of localization length agrees well with other studies in 2D TMDCs [11–14]. Considering a defect density of $n_t \sim 1x10^{13}$ $cm^{-2}$ [14] the average defect distance is $a \sim 3$ $nm$: thus when $\xi \geq a$, the states becomes delocalized (for n ≥ 2x$10^{12}$ $cm^{-2}$) resulting in metal-like transport which corroborates our observation in Fig. 3a.

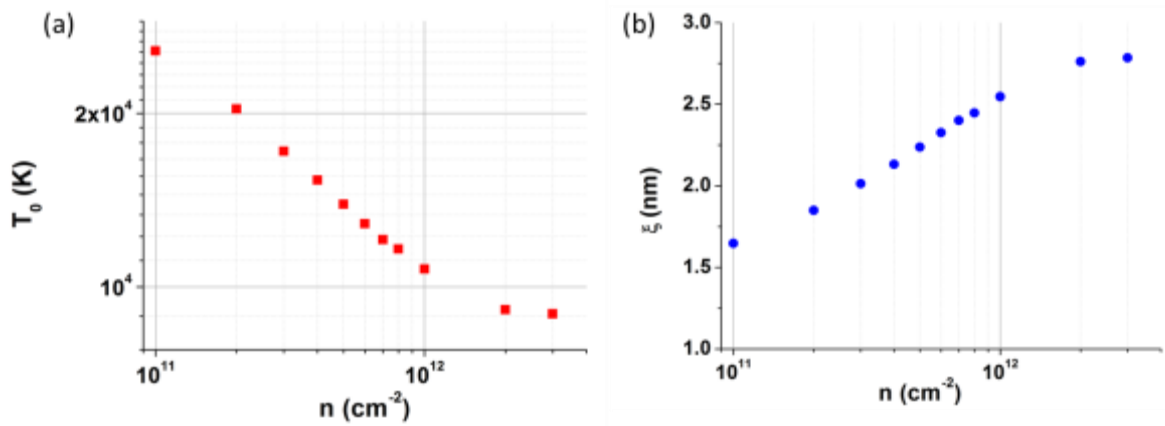

**Figure S5**. (a) Variable-Range Hopping (VRH) correlation energy scale, $T_0$ plotted as a function of the backgate modulated carrier concentration, n for a monolayer $MoS_2$ sample, extracted from Fig. S4(c). The value of $T_0$ agrees well with literature [7,13]. (b) The localization length, $\xi$ extracted from 2D Mott VRH for a fixed density of hopping states at the Fermi Level.

**Seebeck vs Temperature for monolayer and bilayer in the high temperature metallic regime:**

Figure S6 below shows clearly the *S~T* relationship in the metallic phase in the temperature range 300-350 K for a monolayer and a bilayer sample:

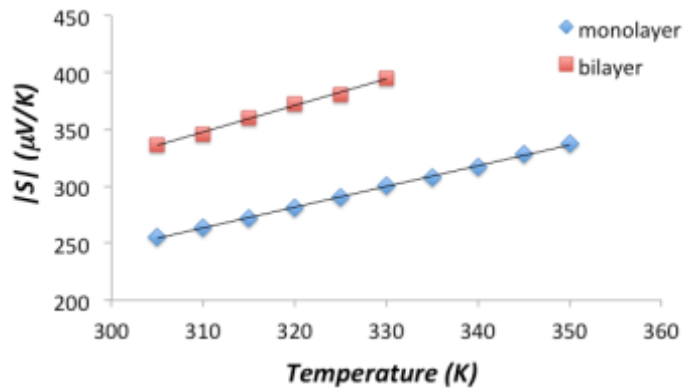

**Figure S6.** Magnitude of Seebeck (note, Seebeck values are negative since both monolayer and bilayer MoS$_2$ are *n*-type as illustrated in the main manuscript) plotted as a function of temperature for monolayer and bilayer samples in the metallic regime showing a linear relationship.

We would need many more temperature points (3 orders of magnitude in terms of the temperature range) in order to extract exact temperature dependence of Seebeck across the MIT transition (such as Fig. 1(b) in Demishev et. al. [15]). For the monolayer sample as discussed in the main manuscript, the highly conductive metallic phase is observed for high carrier concentrations and at temperatures T>280K (Figure 3a).

**Calculation of Seebeck Coefficient and Fermi Level (with respect to the conduction band minimum (CBM):**

In order to calculate the Seebeck coefficient for the monolayer and bilayer samples, the position of the Fermi Level, $E_f$ with respect to the CBM, $E_c$ given by $(E_f - E_c)$ must be known. Given that the doping due to the backgate pushes the 2D MoS$_2$ channels into the degenerate limit (evidenced by the decreasing conductivity with temperature and the linearity of the measured Seebeck as a function of temperature), Fermi-Dirac statistics need to be used. Boltzmann statistics are only valid in the limit that $|E_c$-$E_f| \gg k_BT$, which is not the case in our experiments.

Therefore, in the degenerate limit,

$$n = \int_{E_c}^{\infty} D_{2D}(E) f_{FD}(E) dE \qquad \text{(S1)}$$

where $D_{2D}(E) = \dfrac{g_v g_s m^*}{2\rho \hbar^2}$ are the 2D density of states (DOS) ascertained earlier in the Supplementary Information. Here, $g_v$ and $g_s$ are the valley and spin degeneracies respectively and $m^*$ is the band effective mass obtained from the band structure. A summary of the values for monolayer and bilayer are given in Table 1 in the main manuscript:

Note here that the DOS of bilayer MoS$_2$ is ~4 times larger than monolayer MoS$_2$ due to simultaneously higher degeneracy and higher effective mass at the CBM. Also note that the DOS in the 2D limit is energy-independent.

$f_{FD}(E) = \dfrac{1}{\exp^{(E-E_f)/k_BT} + 1}$ is the Fermi-Dirac distribution.

Let $\varepsilon = (E - E_c)/k_BT$ and $\eta = (E_F - E_c)/k_BT$. Then, equation S1 above gives:

$n_{2D} = N_{c,2D} \displaystyle\int_0^{\infty} f_{FD}(\varepsilon) d\varepsilon$, where $N_{c,2D} = D_{2D} \cdot k_BT$ is the effective density of states in two dimensions. Here, $\displaystyle\int_0^{\infty} f_{FD}(e) de = F_0(\eta)$ is just the 0$^{\text{th}}$-order Fermi Integral, which can be evaluated analytically: $F_0(\eta) = \ln(1 + \exp^{\eta})$.

Therefore, in order to relate the Fermi energy to the carrier density, we use the expression

$(E_F - E_c) = k_BT \cdot \left[ \exp^{\frac{n}{N_c}} - 1 \right]$, where $n$ is determined experimentally in the 2D MoS$_2$ channel as explained in the text. $(E - E_c)$ is plotted for monolayer and bilayer MoS$_2$ given in Figure S7 below:

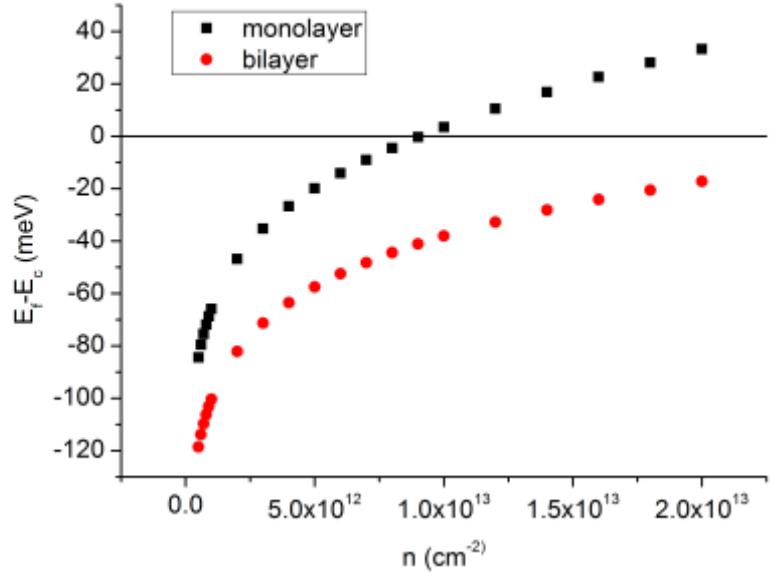

**Figure S7.** The position of the Fermi level ($E_f$) with respect to the CBM ($E_c$) for a monolayer and bilayer sample as a function of back-gate induced carrier concentration at 300K

For the Seebeck Coefficient, the 2D DOS at the CBM are accessible by the window function, $F_{w1}(E,T) = (E-E_F) \times \{-df_{(E,T)}/dE\}$, which has an energy width of ~$2k_BT$. At 300K, this value is ~52meV. Therefore, we would expect that the Seebeck coefficient measured in our experiments will be sensitive to the 2D MoS$_2$ DOS for n ~ 2-4x10$^{12}$ cm$^{-2}$ as seen in Figure S7 above for both monolayer and bilayer MoS$_2$.

The Seebeck coefficient is calculated using the linearized Boltzmann Transport Equation in the relaxation time approximation given in Equation (1) in the main manuscript and reproduced below:

$$S = \frac{1}{qT}\left[\frac{\int_{Ec}^{\infty}\frac{df_{FD}}{dE}D_{2D}(E)(E-E_F)\tau(E)dE}{\int_{Ec}^{\infty}\frac{df_{FD}}{dE}D_{2D}(E)E\tau(E)dE}\right] \qquad (S2)$$

$\tau(E)$ is the energy-dependent scattering or relaxation time. This scattering time is often assumed to have an energy-dependence of the form $\tau = \tau_0 E^r$, where the exponent, r, depends on the dominant scattering mechanism [16]. For scattering of acoustic phonons, it has been shown that $\tau(E)$ scales with the density of states [17], thus, r=0 for acoustic phonon-limited scattering in 2D in the single parabolic band model. For charged impurity scattering, the scattering roughly has the energy dependence $\tau(E) \propto E^{3/2}$, for a simple model for elastic scattering where the bands are assumed to be parabolic and the impurity is screened with a Thomas-Fermi type screening in 2D [17].

Therefore, calculating the Seebeck coefficient as a function of the carrier concentration, *n*, elucidates the dominant scattering mechanism of electrons in the 2D MoS$_2$ channels. Again, by considering $\varepsilon = (E-E_c)/k_BT$ and $\eta = (E_F-E_c)/k_BT$, using the energy-independent DOS, *D$_{2D}$,* and accounting for the energy-dependent scattering rate described above, the equation can be written out as:

$$S = \frac{-k_B}{q}\left[\eta - \frac{(r+2)\int\limits_{\varepsilon=0}^{\infty} f_{FD}\,\varepsilon^{r+1}d\varepsilon}{(r+1)\int\limits_{\varepsilon=0}^{\infty} f_{FD}\,\varepsilon^{r}d\varepsilon}\right] \qquad (S3)$$

,where the integral in S2 is simplified by using integration by parts:

$$\int_0^\infty \frac{df_{FD}}{dE}E^n dE = \left[f_{FD}\cdot E^n\right]_0^\infty - n\int_0^\infty f_{FD}E^{n-1}dE$$

In Equation S3, the integrals are just Fermi Integrals of different half-orders, depending on the value of *r*. We calculate the Seebeck coefficient for two values of r:

(a) *r =0* in the case of scattering dominated by phonons in 2D [16,18,19]:

$$S_{r=0} = \frac{-k_B}{q}\left[\eta - \frac{2}{1}\frac{F_1(\eta)}{F_0(\eta)}\right]$$

(b) *r =3/2* a hypothetical case to show how the energy dependence of the relaxation time can enhance S:

$$S_{r=3/2} = \frac{-k_B}{q}\left[\eta - \frac{7}{5}\frac{F_{5/2}(\eta)}{F_{3/2}(\eta)}\right]$$

Typically, *r=3/2* is the exponent for electrons scattered by charged impurities in three dimensions, but it can be different for two dimensions depending on the approximations used [16,17,20]. These equations are used in the manuscript in Fig. 4c and 4d to compare to the experimentally obtained Seebeck coefficients for both monolayer and bilayer MoS$_2$. Numerical integration was performed using the function fermi.m in Matlab® [21].

Note here that the value of the Seebeck coefficient does not depend on the absolute value of the scattering time, $\tau_0$ (Eq. S2). Hence, while the mobility of the samples measured is limited directly by the scattering time, given by $\mu = e\,\tau/m*$, the Seebeck is only sensitive to the availability of the DOS near the Fermi energy and the energy-dependence of the scattering term.

Comparing the experimentally measured Seebeck coefficient to theory strongly suggests that the scattering is dominated by electron-phonon scattering. The electron-phonon scattering rate in monolayer has been previously calculated from first principles [18,22]. Over an energy range of 50 meV, the scattering rate in both the K and Λ valleys is indeed constant, with a total scattering rate of roughly 1x10$^{13}$ s$^{-1}$ over all phonon modes. However, the mobilities in our samples are lower than the intrinsic phonon-limited mobility of ~410 cm$^2$/V.s [18]. Our measured mobilities are similar to other measured mobilities for MoS$_2$ on SiO$_2$ [6,23], suggesting that substrate-monolayer coupling may significantly alter the phonon channels available to carriers in MoS$_2$.

**Thermoelectric measurement setup and details of technique:**

After the annealing steps as mentioned in Methods, at each device temperature, we first run the transfer curve as described earlier to obtain the Field-Effect Mobility. In order to measure the Seebeck, a series of experiments is performed:

(a) Two resistance thermometers (used to obtain the temperature gradient, $\Delta T$) are calibrated in order to obtain their resistance change as a function of global temperature. This is done by measuring the local 4-probe resistance of the thermometers (using two SRS850 lockin amplifiers) over a series of temperatures around the global temperature. Figure S8 shows a typical measurement. Once the thermometer calibrations are performed, the measured 4probe resistance can be converted to the local temperature at the electrode locations, i.e., $T_{top}$ and $T_{bottom}$.

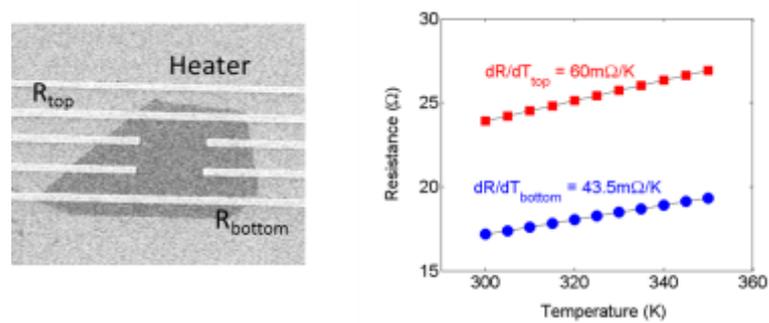

**Figure S8**. Resistance vs Temperature for two thermometer, $R_{top}$ and $R_{bottom}$

(b) Next, under zero gate voltage, the heater line calibration is performed. Here, the resistances of the top and bottom thermometers are measured as a function of the DC current (sourced using a Keithley 2400) running through the heater line. Representative data with fitting is shown below in Figure S9:

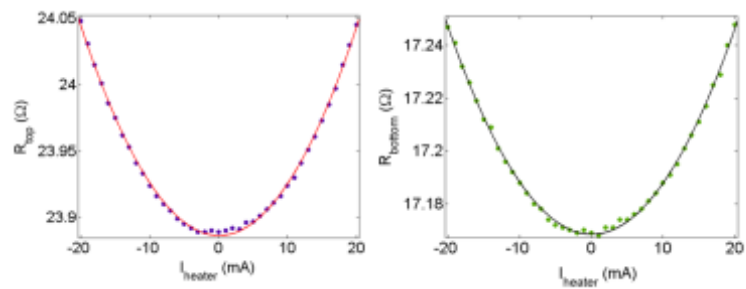

**Figure S9**. $R_{top}$ and $R_{bottom}$ as a function of the heater current, as the heater generates a temperature gradient across the device. As expected, $R_{top} \propto I_{heater}^2$

Using the thermometer calibration in (a), the device temperature gradient can then be accurately measured as a function of the heater current, as illustrated below in Figure S10.

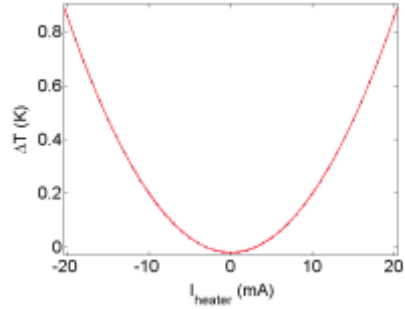

**Figure S10**. Temperature gradient generated across the device measured using the resistance thermometers, as a function of the applied Joule heat from the patterned heater current, $I_{heater}$

Typical noise levels (after averaging 3 runs) are around ~30mK, which is equivalent to the thermal noise of the cryostat (at 300K base temperature). Typical temperature gradients (for a ~5μm length device are $\Delta T$~1K). Hence the noise level in the temperature measurement is ~3%.

(c) Next, the electrical conductivity and open-circuit voltage, $V_{oc}$ (when source-drain current is zero) are measured simultaneously at every gate voltage, $V_g$. Here, three Keithley 2400 sourcemeters are used in conjunction (DC current source for the heater, gate voltage and electrical/seebeck measurement across MoS$_2$ device). Using the $\Delta T$ determined in (b), the Seebeck can be obtained as the slope of the $V_{oc}$ given by $S = -V_{oc}/\Delta T$ as shown in Figure S11. The error in the Seebeck measurement at high gate voltages (in the metal-like MoS$_2$ state) is ~10μV/K, hence typically <5%.

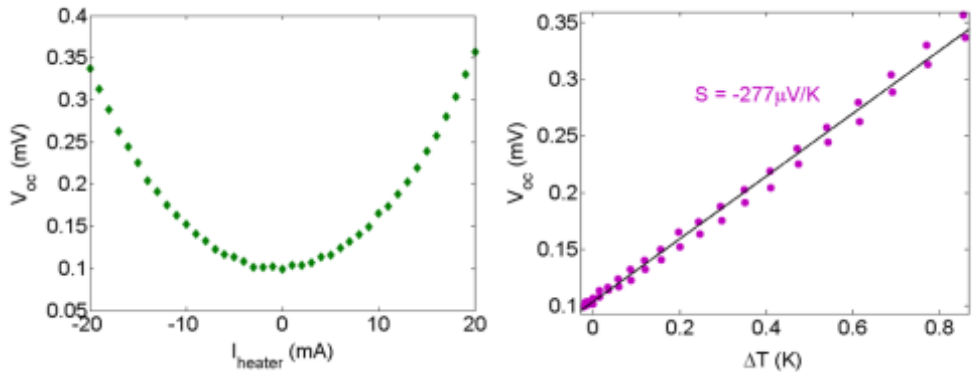

**Figure S11**. Open circuit voltage, $V_{oc}$ measured as a function of the heater current, $I_{heater}$. This is then used to obtain the Seebeck coefficient from the thermometer calibration across the device, given as the slope of $V_{oc}$ vs $\Delta T$.

Note here that the highly doped Silicon wafer (the backgate) acts as a heat sink that controls the temperature gradient across the two electrodes, while the low thermal conductivity SiO$_2$ (gate-dielectric) acts as a thermal barrier between the bottom wafer and the metal electrodes, controlling the actual local temperatures. The heat is generated from the center of the patterned heater and decays linearly on the surface of the SiO$_2$ upon which the MoS$_2$ lies, while the metal electrodes that function as resistance thermometers measure the local temperature gradient in intimate contact with the MoS$_2$ as described in Methods in detail. The heat flows out from the EBL-defined heater isotropically in all directions in the SiO$_2$ substrate. Since the MoS$_2$ is atomically thin, a very small portion of that heat generated by the heater actually flows through the MoS$_2$ cross-section. The key to accurate Seebeck measurement of the $MoS_2$ lies in measuring the local temperature across the $MoS_2$ at the same locations as the open-circuit voltage, which the design is able to accomplish. The high resistance in the OFF state of the $MoS_2$ ($V_g \leq V_t$) introduces additional capacitive coupling and hence the noise levels of the Seebeck measured are higher. We do not measure the Seebeck in the OFF state in this study.

**Temperature dependent contact resistance and monolayer electronic stability:**

Temperature-dependent two-probe (*2p-G*) and four-probe (*4p-G*) conductance of monolayer MoS₂ was measured to extract the contact ratio, $\chi_c$ = G$_{4p}$/G$_{2p}$ at 100K, 150K, 200K, 250K and 300K. The contact ratios of our monolayer MoS₂ at 100 K and 300 K (Figs. S12a and S12b respectively) are comparable to Ti/Au contacts used in literature [24]. At 100 K, the contact ratio drops from 10 (@$V_g$ = −60 V) to 2.5 at high electron concentration ($V_g$=70 V), while at 300 K, the contact ratio remains small ~1 (@$V_g$ = −60 V) to ~2 at high electron concentration ($V_g$=70 V). The two-probe conductance that is measured simultaneously with the Seebeck (*2p-G* (*V$_{oc}$*)) is also different from the two-probe (*2p-G*) and four-probe (*4p-G*) measurements, which indicates the device changes when exposed to air and is reloaded into the cryostat (despite the same *in-situ* annealing of 1 hour at 475 K under high vacuum). Thus, it is imperative to measure the Seebeck and electrical conductivity simultaneously for the device powerfactor as has been done for all devices in the main manuscript. Note that the numerical value of the contact ratio is only used to estimate the correct density of states in the main text, Equations (2) and (3) and not in the reported values of powerfactor.

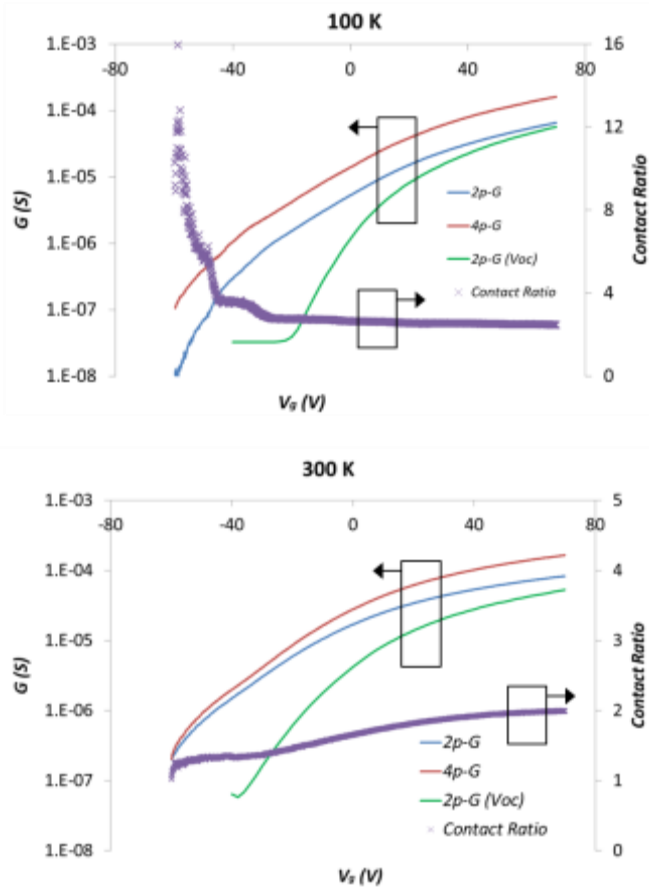

**Figure S12.** (a) Gate-dependent two-probe (blue line) and four-probe (red line) conductance measurements of the monolayer MoS₂ at 100 K. The contact ratio (purple star) is determined by $\chi_c$ = G$_{4p}$/G$_{2p}$; drops from 10 (@$V_g$ = −60 V) to 2.5 at high electron concentration ($V_g$=70 V). (b) Gate-dependent two-probe (blue line) and four-probe (red line) conductance measurements of the monolayer MoS₂ at 300 K. The contact ratio (purple star) is low ~1 (@$V_g$ = −60 V) to 2 at high electron concentration ($V_g$=70 V).

**MoS$_2$ layer number characterization, Atomic Force Microscopy (for layer thickness) and High-Resolution Scanning Electron Microscopy (for roughness, ripples and/or folds):**

The monolayer, bilayer and trilayer MoS$_2$ samples are selected and identified based on characterization from optical contrast, photoluminescence images and Raman spectroscopy. The separation between Raman-active modes A$_{1g}$ and E$^1_{2g}$ decreases as the layer thickness decreases from three layers to one layer, as has been reported in literature. [24–26] The separation of the A$_{1g}$ and E$^1_{2g}$ peaks are 18 cm$^{-1}$ for the monolayer, 22 cm$^{-1}$ for the bilayer and 24 cm$^{-1}$ for the trilayer (Fig. S13a). Moreover, the monolayer MoS$_2$ exhibits strong photoluminescence, due to the direct bandgap nature (Fig. S13b). [27]

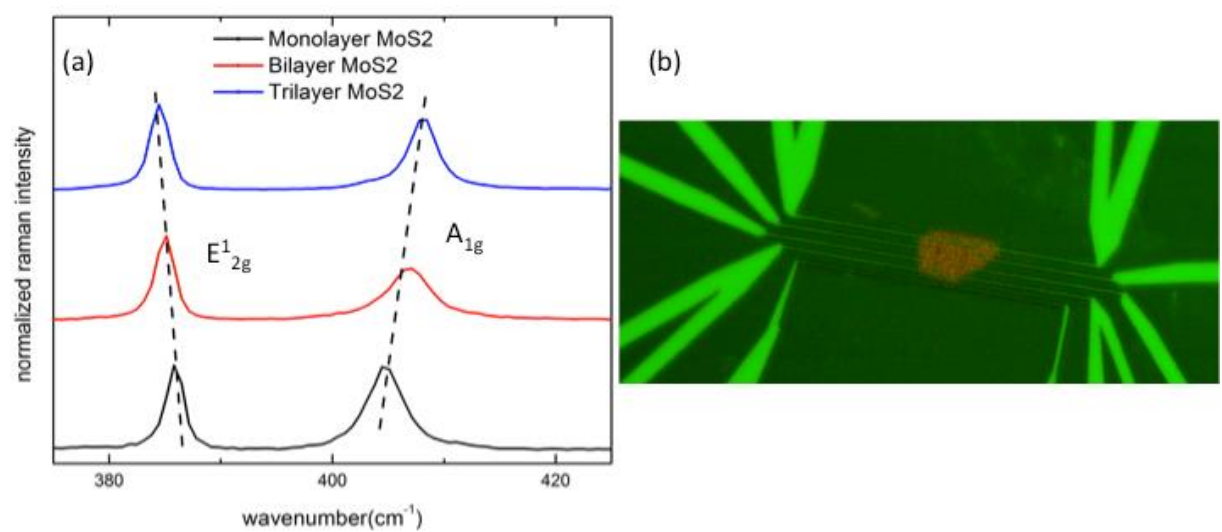

**Figure S13.** (a) Thickness dependence of the Raman spectra of MoS$_2$. The Raman spectrum of MoS$_2$ has two prominent peaks: an in-plane (E$^1_{2g}$) mode and an out-of-plane (A$_{1g}$) mode. As MoS$_2$ becomes monolayer these two modes evolve with thickness. The in-plane mode upshifts to 386 cm$^{-1}$ and the out-of-plane downshifts to 404 cm$^{-1}$. The difference of these two modes (~18 cm$^{-1}$) can be used as a reliable identification for monolayer MoS$_2$. (b) Confocal image of a monolayer MoS$_2$ thermoelectric device. The monolayer MoS$_2$ shows strong photoluminescence, due to the direct bandgap property. The red channel is the photoluminescence channel, indicating the shape of monolayer MoS$_2$. The green channel is the scattering channel of the incident laser, indicating the geometry of the device.

In order to get the real thicknesses for monolayer and bilayer to guarantee the accuracy of estimation of electric conductivity, we also conduct AFM measurements. AFM images (Fig. S14) were acquired from Dimension 3100 scanning probe microscope under non-contact mode using probes with tip radius less than 10 nm. The thickness was measured from the step between the MoS$_2$ crystal and the underlying SiO$_2$ substrate. The scanning parameters have been carefully selected to yield the true thickness due to the difference of tip-sample interaction over the substrate and the crystal [28]. The devices were prepared after following identical fabrication steps and annealing procedures as described in Methods. The devices were annealed in a high vacuum of 5×10$^{-6}$ torr at 475 K for 1 hour, which removes the tape residue, particles and absorbed water molecules from the surface and reduces the surface roughness. Representative monolayer and bilayer samples show a thickness of 0.66 nm for the monolayer (with a rms roughness less than 0.2 nm) and 1.31 nm for the bilayer (also with a rms roughness less than 0.2 nm) respectively (Fig. S14), similar to reported values in literature [4,23,29]. We have included appropriate error bars in the electrical conductivity estimation in the main manuscript in the determination of powerfactors (Fig. 2c in main manuscript).

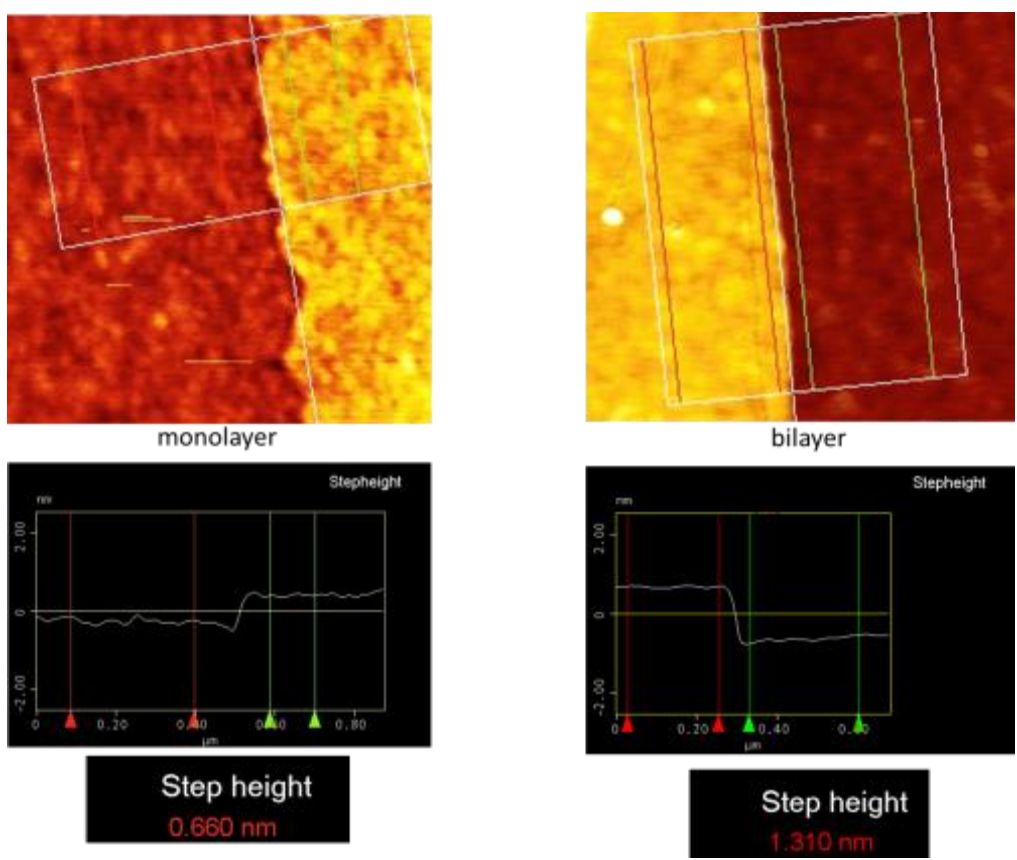

**Figure S14.** AFM measurements of monolayer and bilayer MoS₂. The thickness of monolayer is about 0.66 nm and the thickness of bilayer is about 1.31 nm.

We also discuss below ripples on flakes that might affect the thickness of monolayer and bilayer. Formation of ripples in two-dimensional graphene [30–32] and MoS₂ [33–35] has been discussed in much detail in literature. These 2D crystals form ripples in order to stabilize their structure when suspended. Notably, while monolayer MoS₂ and graphene are exfoliated upon a suspended platform, they have ripples of similar height and periodicity [33]. However, bilayer MoS₂ is shown to be much flatter than monolayer MoS₂ as well as bilayer graphene [33], therefore the likelihood of ripples affecting mobility and hence electron transport is minimal in our bilayer devices. Further, when exfoliated upon a substrate, the presence of an underlying support has been hypothesized to relax the stresses and there is no experimental evidence of ripple formation so far in supported graphene and/or MoS₂ samples [30], unless intentionally induced by using a flexible substrate [35]. A new report shows predominant ripples that are visible in both SEM and AFM for monolayer MoS₂ that is grown (not exfoliated) on SiO₂. However, this is expected for as-grown samples since during the growth process, substrate strain due to lattice mismatch is relaxed by spontaneous ripple formation [34]. In our procedure of exfoliating samples, we rarely notice ripples, usually observable with high resolution SEM as evidenced by images in Fig. S15(a) below. Occasionally, the samples fold over themselves (black arrows), but this is clearly visible either with an optical microscope or under a Scanning Electron Microscope (SEM) and we discard such samples and do not use them for fabricating our thermoelectric devices. Similarly we discard those samples with small ripples (red arrows) or cracks (green arrows) and/or multilayer overlap (blue arrow) as seen in Fig. S15(a) below.

Figure S15(b) shows representative SEM samples that have clean surfaces without ripples/cracks or folds, which are typical of the samples that we use for our thermoelectric devices, and as is also evidenced by clean and smooth surfaces in our AFM pictures shown in Fig. S14.

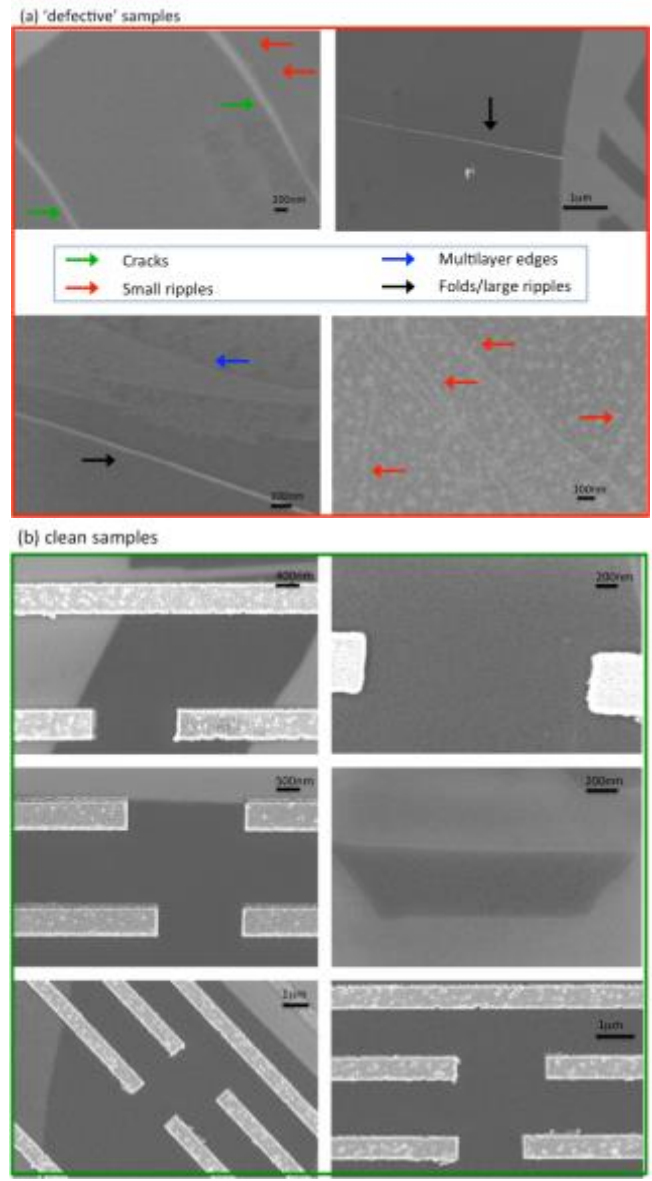

**Figure S15.** (a) High Resolution Scanning Electron Micrographs (HR-SEM images) of samples with folds, cracks, large/small ripples, and multilayer boundaries clearly visible. We classify such samples as 'defective' during Electron Beam Lithography (EBL) patterning and do not use them for our measurements. (b) HR-SEM images of representative clean samples with very low roughness, and showing no evidence of any 'defects' including ripples. The bright lines are electrodes formed after metal deposition.

**Current induced Joule heating, current crowding effects and thermoelectric potentials at the Metal-MoS₂ contacts:**

Since the Seebeck measurements are performed under open circuit conditions such current-induced effects will be zero while measuring the thermopower. Usually these effects are large under high source-drain currents and voltages (typically $I_{ds}$~150 μA/μm and $V_{ds}$~1 V) based on the experiments on graphene [36,37], while measuring electrical conductivity. In our measurements of electrical conductivity, $I_{ds}$<0.1 μA/μm and $V_{ds}$<10 mV, which means that the effects of heating and thermoelectric voltage generation should be orders of magnitude smaller than those reported for graphene.

In order to verify experimentally that the measured electrical properties do not depend on the source-drain current, $I_{ds}$, we performed an experiment where we measured the 2-probe electrical resistivity as a function of a series of atypically high $I_{ds}$ values (~500 times higher than values we typically use for electrical conductivity experiments: see Fig. S16 below). The length of this device is 9 μm, thus the current densities are: 5.56 μA/μm ($I_{ds}$=50 μA), 11.11 μA/μm ($I_{ds}$=100 μA), 22.22 μA/μm ($I_{ds}$=200 μA), and 55.56 μA/μm ($I_{ds}$=500 μA) respectively. The electrical resistivity **does not** depend upon the magnitude of the current and indicates that joule-heating, current-crowding or thermoelectric effects should be negligible even at such high current densities for the measured electrical conductivity.

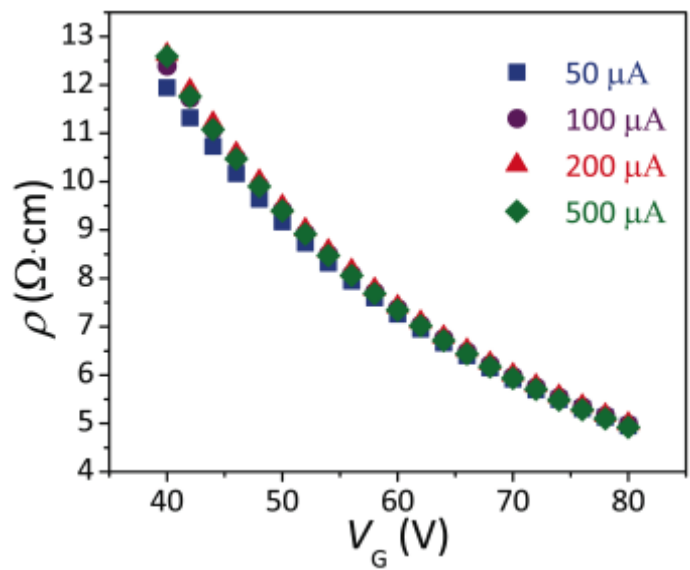

**Figure S16.** Resistivity dependence of source-drain current. For different values of the drain-source current, $I_{ds}$, the measured resistivity is identical, especially at high gate voltages.

We have performed an additional experiment to monitor the temperature rise in the MoS₂ (by measuring the Raman shift of the MoS₂ E₂g peak) while driving an increasing electrically supplied power through the device. The summary of this data is shown in Fig. S17 below. The temperature calibration was performed on the same device in a home-built Raman cryostat where the stage temperature was monitored while the Raman-active E₂g peak location was measured. By calibrating the peak-shift as a function of temperature, we can ascertain the local temperature of the MoS₂ device given by $\Delta T_{MoS_2}$ [38,39]. Since $\Delta T_{MoS_2} \propto \left( I_{ds} V_{ds} \right)$, we see that the temperature rise in MoS₂ only begins at ~2 mW (corresponding to an applied current of ~250 µA).  The device length is 5 µm giving a current density of ~50 µA/µm which is ~500 times higher than the typical values used for measurement of two-probe electrical resistivity (typically <0.1µA/µm) that we have reported in our main manuscript.  This is also consistent with the high current density experiments performed above. Therefore, we would expect current crowding, joule heating and thermoelectric effects to be **negligible** in our measurements; these would only manifest at higher values of current and voltages, which we do not use in our measurements.

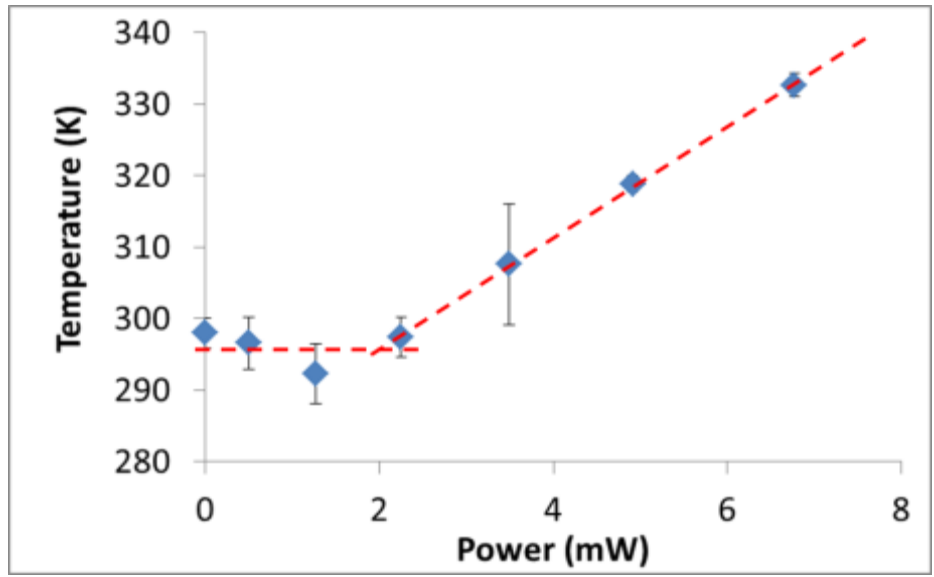

**Figure S17.** Temperature measurement of MoS₂ device using Raman spectroscopy. The temperature rise in MoS₂ only begins at ~2 mW (corresponding to an applied current of ~250 µA).

**Comparison to other thermoelectric materials:**

The gate-modulated Seebeck coefficient of monolayer MoS₂ (shown as $\alpha$ in µV/K in Fig. S18a) is plotted as a function of electrical conductivity (shown as $\ln\sigma$ in $\Omega^{-1}cm^{-1}$) in comparison to traditional thermoelectric materials. Evidently, $\alpha = m(b-\ln\sigma)$ [40], with the slope m $\approx k_B/e$. A larger value of the intercept, b, indicates a larger powerfactor ($\alpha^2\sigma$). The thermoelectric performance of monolayer MoS₂ is comparable with that of high-performance thermoelectric materials such as Bi₂Te₃ and PbTe. Similarly, the monolayer MoS₂ matches Bi₂Te₃ in the $\alpha^2\sigma$-$\sigma$ plot (Fig. S18b), while the bilayer MoS₂ has a higher powerfactor at the same conductivity, indicating superior thermoelectric performance. The higher value of the intercept, b, is linked to the large effective mass, $m*$ and mobility, $\mu$ of the layered MoS₂ [40,41].

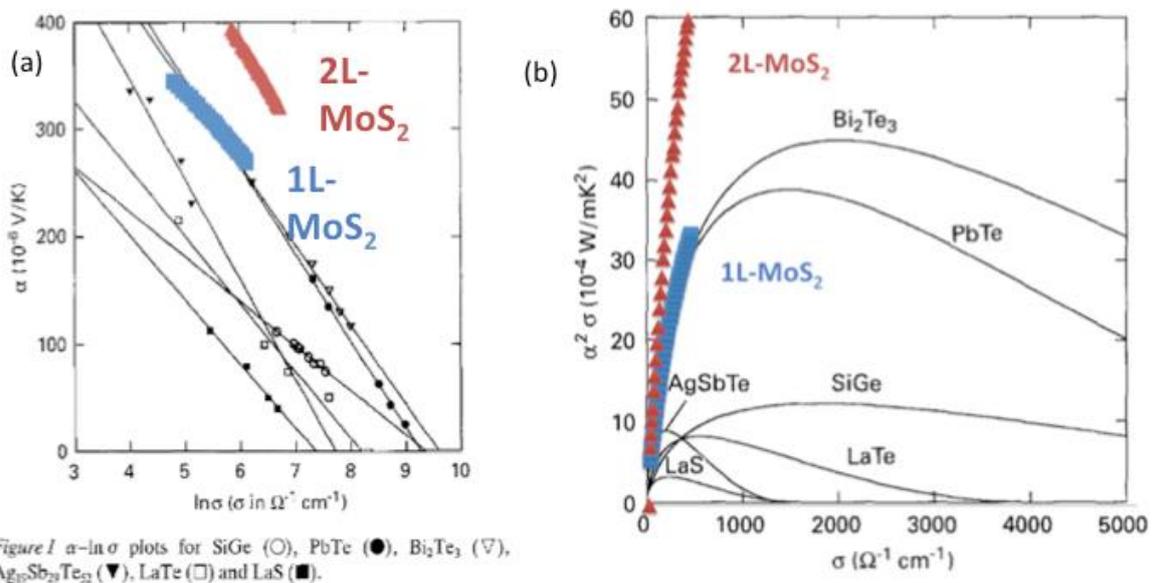

*Figure 1* α-ln σ plots for SiGe (◯), PbTe (●), Bi₂Te₃ (▽), Ag₁₀Sb₂Te₅₂ (▼), LaTe (☐) and LaS (■).

**Figure S18.** (a) $\alpha$-$\ln\sigma$ plot of monolayer and bilayer MoS₂ in comparison of thermoelectric performance with traditional thermoelectric materials, adapted from Rowe et. al. [40] (b) $\alpha^2\sigma$-$\sigma$ plot of monolayer and bilayer MoS₂ in comparison of thermoelectric performance with traditional thermoelectric materials. The thermoelectric performance of monolayer is comparable with that of high-performance thermoelectric material Bi₂Te₃, while the bilayer indicates superior thermoelectric performance.

Additional two monolayer samples (Fig. S19) show saturation of the powerfactor at even higher gate voltages (carrier concentrations), close to 1-1.5×10²⁰ cm⁻³, which is behavior similar to a degenerately doped semiconductor. Hence, in a regime in which band conduction dominates (**high *n* and high *T***) as shown by our phase diagram in the main manuscript Fig. 3a, until we reach very large gate voltages, the powerfactor is expected to monotonically increase. Unfortunately, this regime was not reached for many of the devices measured as the maximum gate voltage applied depends on the gate oxide quality and voltage breakdown (due to the metal contacts or the MoS₂ channel), which are sample-dependent.

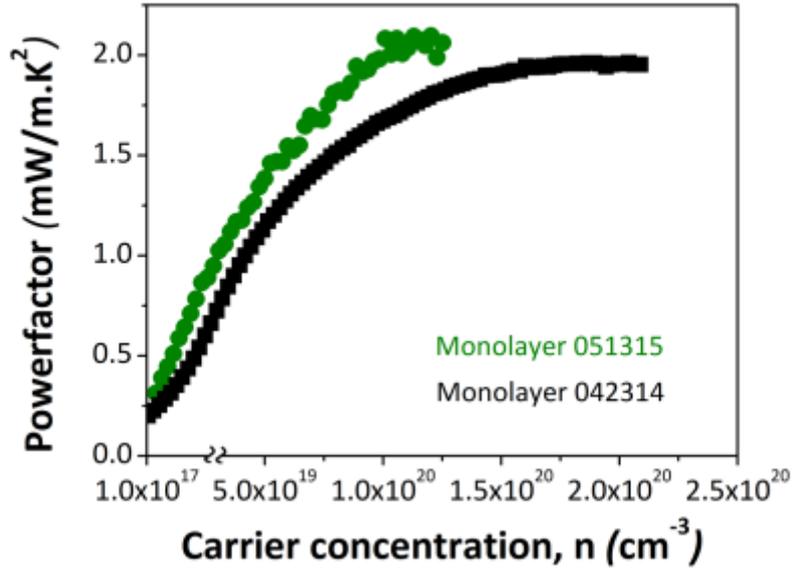

**Figure S19.** Carrier concentration dependence of powerfactor. At high carrier concentration, the powerfactor shows saturation.

The table below summarizes the thermoelectric properties of traditional thermoelectric materials in comparison with our MoS₂ devices (last two rows):

| Material | $\mu$ (cm²/V·s) | carrier concentration (cm⁻³) | $\rho$ (mΩ-cm) | $|S|$ (μV/K) | PF (mW/m·K²) | $m^*/m_0$ |
|---|---|---|---|---|---|---|
| Bi₂Te₃ [Shigetomi, Mori @1956] [42] | $p \sim 280$ <br><br> $n \sim 200$ | $\sim 1 \times 10^{18}$ <br><br> $\sim 1.4 \times 10^{19}$ | $\sim 1.3$ <br><br> $\sim 1.6$ | $\sim 180$ <br><br> $\sim 150$ | $\sim 2.5$ <br><br> $\sim 1.4$ | $\sim 1.26$ <br><br> $\sim 1.07$ |
| Bi₂Te₃ [Harman, Paris, Miller, Goering @1957] [43] | $p \sim 540$ <br><br> $n \sim 400$ | $5 \times 10^{18}$ <br><br> $8 \times 10^{18}$ | $\sim 2$ <br><br> $\sim 2$ | $\sim 150\text{-}200$ (both $n$- and $p$-type) | $\sim 2.0$ | $0.46$ <br><br> $0.32$ |
| Bi₂Te₃ [Satterwaithe, Ure @ 1957] [44] | $p \sim 410$ (ZR) <br><br> $p \sim 430$ (D4) | $2 \times 10^{19}$ <br><br> $3 \times 10^{18}$ | $0.76$ <br><br> $4.84$ | $< 200$ (both $n$- and $p$-type) | $5.2$ <br><br> $0.8$ | $-$ |
| | $p\sim$ 680 (D7) | $4\times10^{18}$ | 2.3 | | 1.7 | |
| | $n\sim$ 330 (D5) | $9\times10^{17}$ | 21.04 | | - | |
| | $n\sim$ 440 (D13) | $3\times10^{17}$ | 47.34 | | - | |
| BiSbTe crystal [Caillat et. al.@1992] [45] | **$n\sim$ 150 (S1)** | **$7\times10^{19}$** | **0.6** | **~180** | 5.4 | ~0.7 |
| | $n\sim$ 200 (S2) | $3\times10^{19}$ | 1 | ~200 | 4.0 | |
| BiSbTe crystal [LaHalle-Gravier, Lenoir, Scherrer & Scherrer @1998] [46] | **$n\sim$ 90** | **$4\times10^{19}$** | **1.6** | **250** | 3.9 | - |
| | **$n\sim$ 50** | **$1.6\times10^{20}$** | **0.8** | **175** | 3.8 | |
| <span style="color:red">2L-MoS$_2$ [this work]</span> | <span style="color:red">$n\sim$ 64</span> | <span style="color:red">$8\times10^{19}$</span> | <span style="color:red">1.36</span> | <span style="color:red">340</span> | <span style="color:red">8.5</span> | <span style="color:red">~8.1</span> |
| <span style="color:red">1L-MoS$_2$ [this work]</span> | <span style="color:red">$n\sim$ 37</span> | <span style="color:red">$1.2\times10^{20}$</span> | <span style="color:red">2.65</span> | <span style="color:red">283</span> | <span style="color:red">3.0</span> | <span style="color:red">~2.1</span> |

The mobilities of commercial thermoelectric materials are higher at lower doping concentrations. Our equivalent 3D bulk carrier concentration (based on the thickness of ~1.3 nm for bilayer) is higher ~1×20 cm$^{-3}$ and the mobilities are comparable to values for Bi$_2$Te$_3$ and BiSbTe, and as expected mobility drops with an increase in carrier concentration. At these high electron densities, we observe a larger Seebeck value due to the high valley degeneracy and large effective mass, which explains why our values of powerfactor for 2D MoS$_2$ are about the same or larger than traditional thermoelectric materials.